# Free-electron interactions with photonic GKP states: universal control and quantum error correction

Gefen Baranes[1,2,4†], Shiran Even-Haim[1,4†], Ron Ruimy[1,4†], Alexey Gorlach[1,4], Raphael Dahan[1,4], Asaf A. Diringer[3,4], Shay Hacohen-Gourgy[3,4], Ido Kaminer[1,4*]

[1]*Solid State Institute, Technion-Israel Institute of Technology, Haifa 32000, Israel*
[2]*Department of Physics, Massachusetts Institute of Technology, Cambridge, Massachusetts 02139, USA*
[3]*Faculty of Physics, Technion-Israel Institute of Technology, Haifa 32000, Israel*
[4]*The Helen Diller Quantum Center, Technion-Israel Institute of Technology, Haifa 32000, Israel*

[†]*equal contributors*   [*]*kaminer@technion.ac.il*

**We show that the coherent interaction between free electrons and photons can be used for universal control of continuous-variable photonic quantum states in the form of Gottesman-Kitaev-Preskill (GKP) qubits. Specifically, we find that electron energy combs enable non-destructive measurements of the photonic state and can induce arbitrary gates. Moreover, a single electron interacting with multiple photonic modes can create highly entangled states such as Greenberger–Horne–Zeilinger states and cluster states of GKPs.**

# 1. Introduction

Quantum error correction is essential for reaching large-scale quantum computation. One prominent approach toward this goal is to encode qubit information on continuous variables [1,2] in quantum harmonic oscillators, known as bosonic codes. These codes, and most prominently the Gottesman-Kitaev-Preskill (GKP) code [1], facilitate quantum error correction for fault-tolerant quantum computation [3]. The generation and manipulation of GKP states is a formidable challenge, as it necessitates non-Gaussian operations that typically require strong nonlinearities.

Creating the required nonlinearity can rely on a wide range of physical mechanisms. The nonlinearity can arise from intrinsically non-quadratic Hamiltonians that can be realized in the optical regime using the Kerr effect [4,5] or post-selection by number-resolving photonic measurements [6–8]. The GKP states can also be deterministically generated from cat states [9,10], which, however, still require nonlinearity for their generation. Such nonlinearities are typically counterproductive to the stabilization of GKP states since they increase decoherence by coupling to external degreed-of-freedom, even more so given that such states rely on a large average photon number.

Leading approaches for generating and manipulating GKP states rely on the coupling to matter ancilla qubits, which provide the necessary strong nonlinearity. Such a scheme was demonstrated experimentally with the vibrational motion of trapped ions [11,12], with cavity photons at microwave frequencies coupled to superconducting qubits in circuit QED [13]. A similar ancilla-based scheme was also recently suggested theoretically in optical frequencies using cavity QED [14].

Here we propose a different physical mechanism that provides the needed nonlinear interaction using free electrons that act as ancilla qubits. We show how the fundamental coherent interaction of free electrons and photons, perhaps the most basic interaction in QED, can provide the building blocks for universal quantum computing with GKP states. The interaction provides the strong nonlinearity needed for quantum error correction and universal control of GKP states. This interaction can be used in gate-based [15] and measurement-based [16] computational protocols.

The first step in this direction has recently shown the free-electron-based generation of GKP states [17]. We now unveil the complete picture, describing how the fundamental electron-photon interaction can provide universality and error correction. The underlying interaction can be described as a conditional displacement operator ($CD$) in the joint electron-photons Hilbert space. We follow ideas from circuit QED [13] and use this operator as a building block to create any arbitrary unitary operation in the combined Hilbert space [18].

The idea to use free electrons in the context of quantum optics is inspired by recent advances in ultrafast electron microscopy. Specifically, our work relies on the inelastic scattering of free electrons by electromagnetic fields, which was famously observed in photon-induced near-field electron microscopy (PINEM) [19–26]. This nonlinear scattering provides the additional degrees of freedom required to encode quantum information on the individual electron by coherent modulation of its wavefunction [27–29]. The ability to control the modulated electrons has been studied extensively in theory (e.g., [30,31]) and experiments (e.g., [32–34]). The interaction of such modulated electrons enables photon addition and subtraction [35], measurement of light statistics [36], coherent control of two-level systems [37–42], and generation of entanglement [43]. The same underlying theory enables heralded generation of Fock states of one or more photons [44–47]. Such ideas and experimental achievements support the feasibility of the scheme we propose here.

The use of free electrons as matter ancilla qubits is intriguing for a few practical reasons. Free electrons are versatile in their energy spectrum and can access large range of frequencies, including the optical (and potentially higher) range. This versatility enables to transfer to the optical regime concepts that were only demonstrated in the microwave regime, such as nonlinear ancilla qubits – potentially bypassing inherent technical limitations of scalability and low-temperature operation.

Moreover, the free electrons are fundamentally different from previously purposed matter ancilla qubits because they are *flying qubits*, meaning that only couple temporarily to the photonic mode before they continue propagating. The limited interaction time reduces the decoherence of the photonic mode by its coupling to the ancilla. This coupling decoherence can be characterized by multiple noisy channels, such as inverse-Purcell decay [48] and self-Kerr nonlinearities [49], which pose a stronger limitation for GKP states due to their larger photon number. These decoherence channels are reduced by the short interaction time of the flying electron qubit.

Another advantage provided by the electrons being flying qubits is that they naturally facilitate coupling between spatially separated photonic modes, which enable the generation of multipartite highly entangled states such Greenberger-Horne-Zeilinger (GHZ) states [50] and cluster states [16], important resources for quantum computation and communication [51–53]. These possibilities are presented in our work below.

## 2. Free electrons as ancillas for conditional displacement on photonic states

We define the electron coherent energy comb as a superposition of electron energy states with a Gaussian envelope around a central energy $E_0$,

$$|\text{comb}_{\sigma,\phi}^{\omega}\rangle \propto \sum_n e^{-\frac{n^2}{2\sigma^2}} e^{i\phi n} |E_0 + n\hbar\omega\rangle. \tag{1}$$

Here $|E_0\rangle$ is the state of an electron with narrow (compared to $\hbar\omega$) energy distribution around the energy $E_0$, $\omega$ is the modulating laser frequency, $\sigma$ is dimensionless and shows the effective number of energy states in the electron comb, and $\phi$ is the electron phase controlled by the laser phase. In this paper, we consider the limit of $\sigma \gg \hbar\omega$, and omit the $\sigma$ in the electron comb notation. In this case, the electron comb becomes an approximate eigenstate of the energy displacement operators $b_\omega$, $b_\omega^\dagger$ (satisfying $b_\omega b_\omega^\dagger = b_\omega^\dagger b_\omega = 1$). These operators describe a translation of $\hbar\omega$ in the electron's energy, which corresponds to the emission or absorption of a single photon [54], respectively.

The electron comb can be described as a qubit with the following basis:

$$|0\rangle_e = |\text{comb}_{\phi=0}^{2\omega}\rangle, \quad |1\rangle_e = b_\omega |\text{comb}_{\phi=0}^{2\omega}\rangle. \tag{2}$$

We denote $|\psi\rangle_e = \alpha|0\rangle_e + \beta|1\rangle_e$ as a general free-electron qubit state. The $|0\rangle_e$ state can be generated via a typical electron comb generation scheme [30,31] using a modulation laser with frequency $2\omega$. Universal single-qubit gates [27] over such free-electron qubit states are achievable by multiple PINEM interactions separated by free-space propagation, i.e., drift. Free-space propagation over an appropriate distance corresponds to a rotation around the Z axis on the Bloch sphere, and PINEM interaction corresponds to a rotation around the X axis on the Bloch sphere [27]. See Fig. 1c. and Appendix A5. Coming back to the analogy of coherent light, if we consider the energy translation operator $b_\omega$, then the electron qubit states are eigenstates of $b_\omega^2$ and satisfy $\langle i|_e b_\omega |i\rangle_e \approx 0$ with $i = 0,1$, similar to ladder operators acting on optical cat states. This observation creates an analogy between the creation of GKP states [17] and cat breeding protocols [9].

To describe the interaction of such modulated electrons with quantum photonic states, we quantize the electromagnetic field, as was presented theoretically in [54,55] and was in part demonstrated experimentally in [36]. This interaction can be described using the following scattering matrix:

$$S(g_Q) = D(g_Q b_\omega) = e^{g_Q b_\omega a^\dagger - g_Q^* b_\omega^\dagger a}. \tag{3}$$

Here $g_Q$ is the coupling between the free electron and the photonic mode; its amplitude $|g_Q|$ is controlled by the distance between the free electron and the mode [47] and its phase $\angle g_Q$ by the modulating laser phase. $a, a^\dagger$ are the annihilation and creation operators for the photonic mode. $b_\omega, b_\omega^\dagger$, unlike the photonic operators, commute $[b_\omega, b_\omega^\dagger] = 0$. $D(\alpha) = \exp(\alpha a^\dagger - \alpha^* a)$ is a coherent displacement operator [56].

For the free-electron qubit, $b_\omega = b_\omega^\dagger = \sigma_x$ (see Fig. 1b). The scattering matrix in Eq. (3) is then reduced to a conditional displacement ($CD$) operator, controlled in the X basis:

$$S(g_Q) = D(g_Q \sigma_x) = |+\rangle_e \langle +|_e \otimes D(g_Q) + |-\rangle_e \langle -|_e \otimes D(-g_Q) =$$
$$= \frac{1}{2}\left( \left(D(g_Q) + D(-g_Q)\right) I + \left(D(g_Q) - D(-g_Q)\right) \sigma_x \right) = CD(g_Q). \tag{4}$$

The following chapters show how the free-electron qubit can be used as an ancilla qubit in manipulating GKP states in a wide range of frequencies, including the optical range.

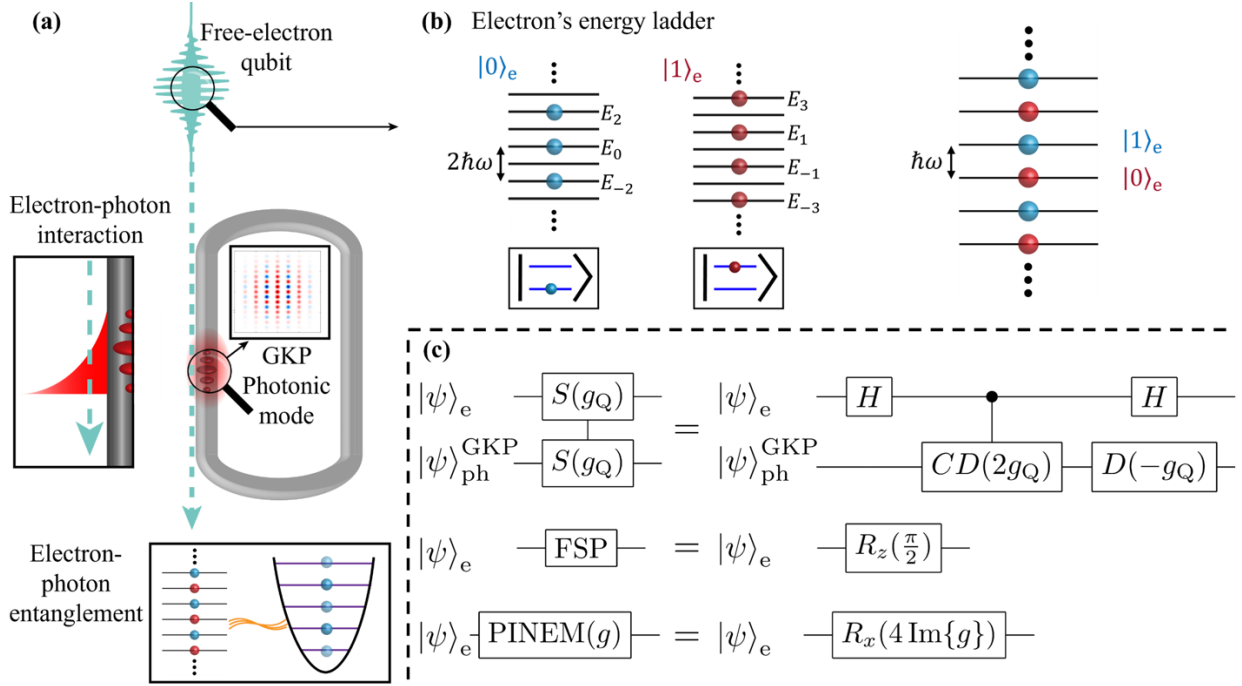

**Figure 1. The free-electron-photon interaction as a fundamental building block for quantum information processing.** (a) The free electron is pre-shaped into a free-electron qubit state (e.g., using laser interactions [30,31]), which interacts with the photonic mode through a near-field coupling. The photonic mode contains a GKP state. The interaction entangles the electron with the GKP state. (b) The free-electron qubit states are shown by their energy spectra, as the even (blue, qubit $|0\rangle_e$) and odd (red, qubit $|1\rangle_e$) comb electrons with $2\hbar\omega$ energy spacing. (c) Building blocks for universal quantum computation on the free-electron qubit [27] and the GKP state (left) and their corresponding circuits (right). The first is the interaction scattering matrix, the second is the free space propagation (FSP) operation on the electron, and the last is the PINEM operation on the electron describing interaction with classical coherent light.

## 3. Universal single-qubit gates and quantum error correction with free-electron ancillas

We focus on the case where the photonic mode is an ideal GKP state [1]. GKP states form a lattice in their Wigner representation [57] and can be defined by the lattice constants $a_{x,y,z}$ (see Appendix B). Stabilization of the GKP code can be achieved with the $CD$ operator from Eq. (4). To create the stabilizers, $g_Q$ should be chosen as a lattice constant $g_Q = \pm a_i$. Pauli gates on the GKP qubit can be achieved in two ways. The first is using the same interaction with $g_Q = \pm\frac{a_i}{2}$, for $i = x, y, z$. The second is by deterministically displacing the photonic state by inserting a laser interaction with the GKP qubit by a beam-splitter. The Hadamard ($H$) gate on the GKP state can be achieved by a $\pi/2$ phase shift of all $g_Q$'s of all the following computation steps [58]. Such an operation can be performed by digitally delaying the electrons. These choices are analogous to the case of regular $CD$ operations based on qubit ancillas [13,58].

When $g_Q = a_i/4$, Eq. (4) gives a controlled Pauli gate $\sigma_i$ on the GKP state, controlled by the electron's state in the X basis. For a non-ideal GKP state, the added displacement $D(-a_i/4)$ needs to be corrected (in post-processing). As an example, the CNOT gate between the free electron and the GKP state is given by $g_Q = a_x/4$:

$$\text{CNOT}_{e\to ph} = \big(H_e \otimes D(-a_x/4)\big)S(a_x/4)(H_e \otimes I). \tag{5}$$

Controlled Pauli gates give the ability to create maximum entanglement between the electron qubit and the GKP state. Moreover, controlled Pauli gates can be used to read out the GKP state by measuring the electron's energy as an ancilla [15] (Fig. 2a).

The $CD$ operator and rotation gates on the ancilla qubit can be used to implement a universal set of gates on the GKP state with an additional feedforward mechanism. In the feedforward mechanism, the next operation is done according to the electron's measurement result.

Rotation gates around $i = x, y, z$ axis with angle $\phi$, $R_i(\phi)$, are achieved with teleported gates by an ancilla qubit [58,59], as shown in Fig. 2c. The initial state of the electron is $|0\rangle_e$. The electron interacts with the GKP state with $g_Q = \frac{a_i}{4}$, $i = x, y, z$ according to the rotation axis and is then measured in the $|\phi_\pm\rangle_e = 1/\sqrt{2}(e^{i\phi/2}|0\rangle_e \pm e^{-i\phi/2}|1\rangle_e)$ basis. The ability to coherently control the electron's qubit state [27] allows measuring it in any desired basis, with additional drift and PINEM interactions for the post-interaction electron. If the measurement result is $|\phi_-\rangle_e$, the Pauli gate $\sigma_i$ is applied to the GKP state, and if the measurement result is $|\phi_+\rangle_e$ there is no need to apply any gate. See Appendix C5 for details on measuring in the $|\phi_\pm\rangle_e$ basis. The $S$ and $T$ gates can be achieved by rotations around the Z axis, with the angles $\frac{\pi}{2}$ and $\frac{\pi}{4}$, respectively.

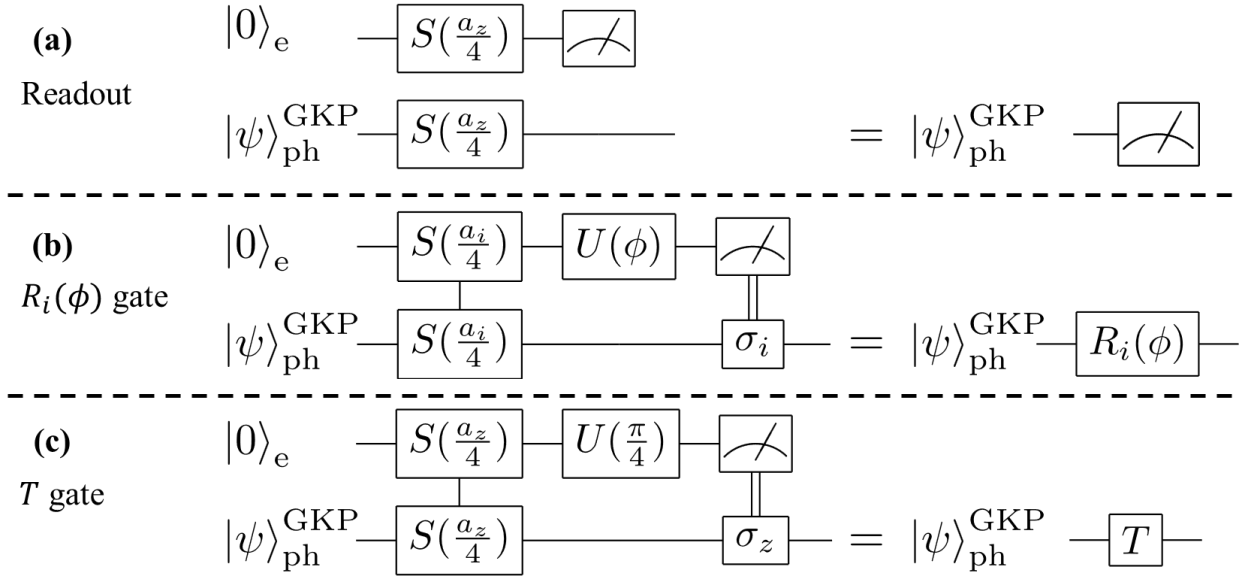

Figure 2. Single-qubit gates induced by the free-electron ancilla. (a) Readout operation: using electron ancilla qubit with interaction $g_Q = \frac{a_z}{4}$, followed by a measurement of the electron to extract the GKP state. Using different axes can be used for readout with any Pauli operator. (b) Rotation gate $R_i(\phi)$ in the $i = x, y, z$ direction: using free-electron ancilla with $g_Q = \frac{a_i}{2}$ performs the gate controlled-Pauli ($C\sigma_i$) on the GKP state, with the electron being the control qubit. Then the electron is measured in the basis $|\phi_\pm\rangle = \frac{1}{\sqrt{2}}\left(e^{\frac{i\phi}{2}}|0\rangle_e \pm e^{-\frac{i\phi}{2}}|1\rangle_e\right)$, using the unitary $U(\phi)$. For feedforward, if the measurement result is $|\phi_-\rangle$, the Pauli $\sigma_i$ gate is applied. (c) $T$ gate: example of rotation gate with $\phi = \frac{\pi}{4}$ and $i = z$.

| Operation | $g_Q$ | Electron state | Feedforward |
|---|---|---|---|
| | | | |
| **Pauli gates $\sigma_i$** | $a_i/2$ | $\|+\rangle_e$ | no |
| **Readout in $i$ basis** | $a_i/4$ | $\|0\rangle_e$ | no |
| **Rotation $R_i(\phi)$** | $a_i/4$ | $\|0\rangle_e$ | If $\|\phi_+\rangle_e$ is measured - none |
| | | | If $\|\phi_-\rangle_e$ is measured - $\sigma_i$ gate |
| **CNOT$_{ph1\to ph2}$** | $g_{Q,1} = \dfrac{a_z}{4}, g_{Q,2} = \dfrac{a_x}{4}$ | $\|0\rangle_e$ | If $\|0\rangle_e$ is measured - none |
| | | | If $\|1\rangle_e$ is measured - $\sigma_z$ gate |

Table 1. **Operations on the photonic state created by a free-electron ancilla for universal quantum computation. Row 1** describes the coupling constant and electron state needed for creating Pauli gates $\sigma_i$ on the GKP state. **Row 2** describes how to use the electron qubit for the readout of the GKP state. **Row 3** is the rotation gate $R_i$ by angle $\phi$, created using a teleported gate with feedforward. **Row 4** shows how to use two electron qubits to create a $CNOT$ gate between two GKP states in different photonic modes.

Altogether, the free-electron qubit ancilla enables the operation of unitary gates, stabilizers, and readouts on the GKP states. These building blocks enable universal quantum computation and error correction [13] using the fundamental free-electron-photon interaction, which can be implemented and controlled in ultrafast electron microscopes [19,22,31,33,36,60].

## 4. The free electron as a flying qubit: creation of GHZ and cluster states

The unique property of a free-electron ancilla as a flying qubit is that a single electron can be used for entangling multiple GKP states. The protocol for a CNOT$_{ph1\to ph2}$ gate between two GKP states in two separated photonic modes is described in Fig. 3a, where one electron qubit interacts with two GKP states. The electron starts in the state $|0\rangle_e$ and interacts with the first GKP state with $g_Q = \frac{a_z}{4}$, then changes the basis using a Hadamard gate ($H_e$) on the electron (see Appendix A5), and then interacts with the second GKP state with $g_Q = \frac{a_x}{4}$. The last step of the protocol for CNOT$_{ph1\to ph2}$ uses feedforward: the electron is measured, and if the measurement result is $|0\rangle_e$, then nothing is applied; but if the measurement result is $|1\rangle_e$, then a Pauli $\sigma_z$ gate is applied to one of the GKP states. The CNOT$_{ph1\to ph2}$ and the universal set of one qubit gates shown in the previous chapter are sufficient for universal quantum computing [15].

The maximally entangled GHZ state can be produced using one electron qubit interacting with multiple photonic GKP states. Each interaction is a CNOT$_{e\to ph}$, which can be implemented with $g_Q = a_x/4$, as presented in Eq. (5). In the final step of creating the GHZ state, the Hadamard gate is applied to the electron. The electron is then measured to disentangle it from the GKP states. Ultimately, a $D(-a_x/4)$ correction should be applied by using one $|+\rangle_e$ electron with $g_Q = -\frac{a_x}{4}$ interacting with all the GKP states to displace it back to the center of the phase space. The procedure is shown in Fig. 3b. This scheme of GHZ states creation can be realized using photonic cavities or a waveguide, as shown in Fig. 3c. For the cavities approach, the distance between the cavities should be designed such that the electron will be phase matched with all modes. For the waveguide approach, the distance between the interaction points must match the electron's path such that the electron will interact with the GKP states. Also, the GKP states must be all phase-matched to the electron [62].

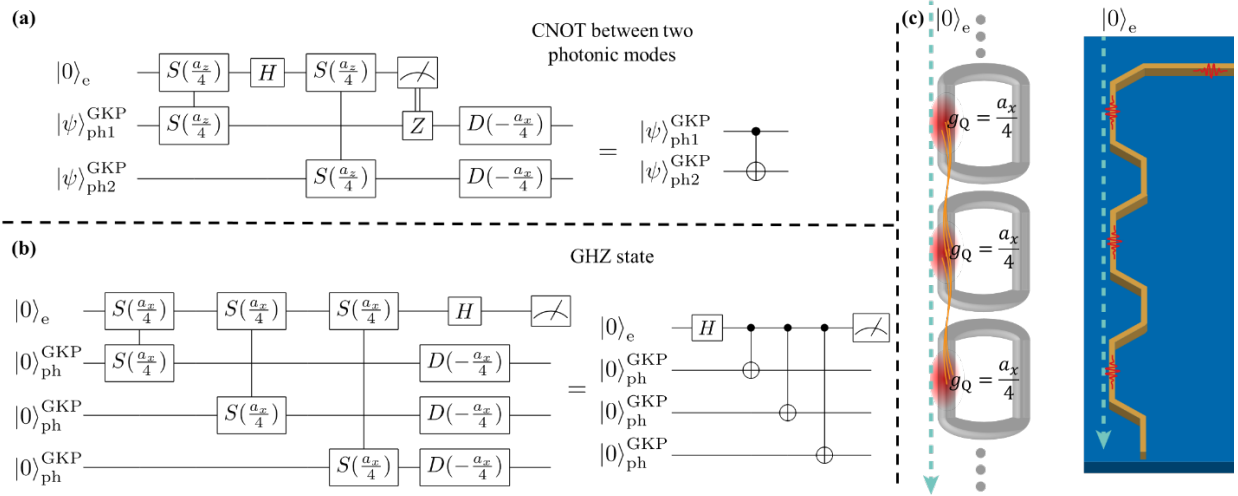

Figure 3. Creation of multiqubit entanglement using free electrons. (a) $\text{CNOT}_{\text{ph1}\to\text{ph2}}$ gate between two GKP states. (b) The scheme for generating a GHZ state of three GKP states. (c) Two approaches for implementing the GHZ state: stationary GKP states in cavities (left) and propagating GKP states in a waveguide (right).

The prospects of free-electron flying qubits include the potential to create the cluster states needed for measurement-based quantum computation schemes. In recent years, much effort has been invested in measurement-based photonic quantum computation, specifically in the optical range. Such schemes require the efficient generation of photonic states and their entanglement into cluster states [16]. Clusters of GKP states [61] are especially desirable because GKP states are robust against photon loss errors [1], and can be easily measured in a different basis with the same operation, as shown in Fig. 2.

Following [63], we can add appropriate propagation distances between the subsequent interactions in the GHZ-creation scheme (shown in the previous section) to add single qubit rotation on the electron and create a 1D cluster of GKP states using a single electron (Fig. 4a). Additionally, combining multiple electron channels can create 2D and potentially higher dimensional cluster states, as shown in Fig. 4b. These higher dimensional schemes are based on the protocol presented in [64] (further discussed in Appendix D2.2). Consequently, free-electron interactions can be used as a building block in measurement-based photonic quantum computation schemes.

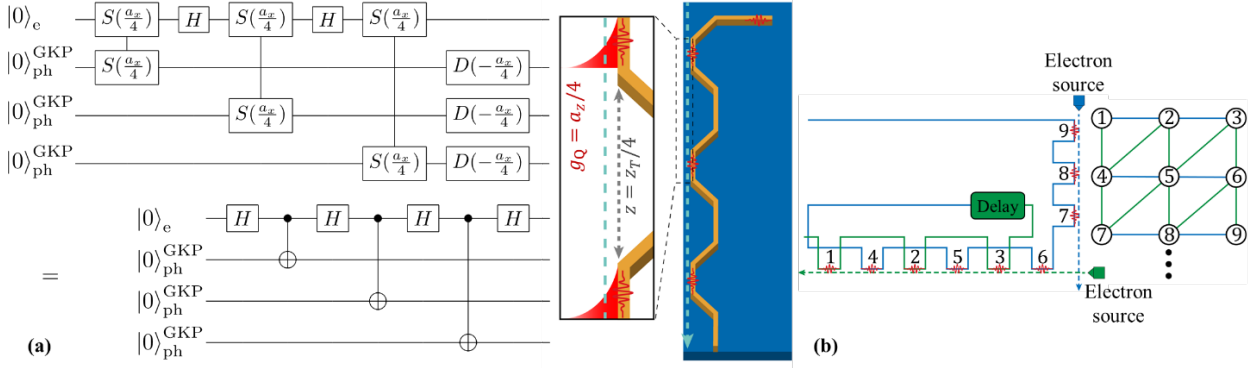

**Figure 4. Using flying qubits for the generation of cluster states. (a)** A scheme for generating a 1D cluster state of GKP states in a photonic waveguide. Quantum circuit description of the proposed scheme (left) and a possible physical scheme using propagating GKP states in a waveguide (right). **(b)** Generation of 2D cluster states of GKP states. A possible implementation using a waveguide, two free-electron sources, and a delay (left). Visualization of the resulting 2D cluster state (right).

## 5. Discussion and outlook

In summary, this paper demonstrates how the coherent interaction between free electrons and GKP states enables projective measurements and universal control over the GKP states. This paper also demonstrates how the interaction of multiple GKP states with the same electron enables the creation of highly entangled states such as GHZ and cluster states. The key to these possibilities is the creation of $CD$ based on the electron interaction. The electron-photon interaction thus reproduces other protocols for GKP state generation in superconducting qubits [13] and ion traps [12]. Going beyond these demonstrations, the free-electron implementation provides additional degrees of freedom to the interaction due to the intrinsic nature of the free electron as a flying qubit.

Although the free electron ancilla qubit provides similar abilities to the circuit QED, trapped ions, and cavity QED counterparts [11-14], the most significant difference between them is that the free electron ancilla is a flying qubit. This allows for high connectivity between the electron and multiple *spatially separated* photonic modes. This fact opens options unavailable in other systems, such as the generation of highly entangled GHZ and cluster states with only one ancilla electron (rather than multiple ancilla qubits [65,66], or multimode coupling [67], which further limits the coherence times and exponentially complicates the physical realization). The flying qubit nature of the electron also implies that it interacts with the GKP state only for a short time (typically ps time scales [33,36]). Therefore, the coherence time of the photonic qubit is not significantly reduced by the free-electron interaction (unlike the case of interaction with ancilla qubits in circuit QED [68]).

It is also interesting to compare the interaction of free-electron qubits with GKP states to other schemes that can be realized in the optical range, such as the beam-splitter interaction of optical cat-states [10,17]. A significant difference is that the creation of free-electron qubits bypasses the need for nonlinear components, as opposed to the creation of cat states. Another advantage of free-electron-based schemes compared to optical ones arise from developments in fast electron counting detectors (direct detection schemes) [69]. Since free electrons are energetic particles, it is easier to achieve number-resolved electron detection than a similar detection with photons.

Looking forward, the free-electron qubit schemes we presented can be generalized to multi-level qudits by changing the electron comb energy gap from $2\hbar\omega$ to $N \cdot \hbar\omega$, where $N$ is an integer corresponding to the number of desired levels [17,29]. Such electron states will be analogous to $N$-legged cat states, which are extremely difficult to generate in optics and can provide additional degrees of freedom that can be exploited for generating and controlling GKP states. This research direction facilitates the tunability of free electrons to provide degrees of freedom that fundamentally differ from their circuit QED or trapped-ions counterparts.

**ACKNOWLEDGMENTS**

We thank Eyal Finkelshtein for the valuable conversations and insights. The research was supported by the European Research Council (ERC Starting Grant 851780-NanoEP) and the European Union Horizon 2020 Research and Innovation Program (under grant agreement No. 964591 SMART-electron).  G.B. would like to acknowledge the support of the Technion Excellence Program for undergraduate students. S.E.H acknowledge the support of the Helen Diller Quantum Center at the Technion. I.K. and A.G acknowledge the support of the Azrieli Fellowship. R.D. would like to acknowledge the support of the Council for Higher Education Support Program for Outstanding Ph.D. Candidates in Quantum Science and Technology in Research Universities. This research was supported by the Technion Helen Diller Quantum Center.